\documentclass[twocolumn, 5p]{elsarticle}
\usepackage{tikz}
\usepackage[shortlabels]{enumitem}
\usepackage{amsmath}
\usepackage{subcaption}
\usepackage{multirow}
\newcommand{\todo}[1]{{\color{red} [TODO] #1}}

\newcommand{\mf}[1]{{\color{red} [MF] #1}}

\usepackage{listings}
\definecolor{mygreen}{rgb}{0,0.6,0}
\definecolor{mygray}{rgb}{0.5,0.5,0.5}
\definecolor{mymauve}{rgb}{0.58,0,0.82}
\lstdefinestyle{example}{
  float=tp,
  floatplacement=tbp,
  abovecaptionskip=-5pt,
  numberstyle=\fontsize{7}{9}\selectfont\ttfamily\bfseries,
  numbers=left,
  numbersep=8pt, 
  xleftmargin=2em,
  frame=tb,
  framexleftmargin=1.5em,
  language=C, 
  basicstyle=\fontsize{7}{9}\selectfont\ttfamily,
  breaklines=true
}

\usepackage{filecontents}
\usepackage[utf8]{inputenc}
\title{COREC: Concurrent Non-Blocking Single-Queue Receive Driver for Low Latency Networking}
\author{Marco Faltelli$^{1,2}$, Giacomo Belocchi$^{1}$, Francesco Quaglia$^{1,2}$, Giuseppe Bianchi$^{1,2}$
\\
$^1$University of Rome Tor Vergata, Italy.\\
$^2$Consorzio Nazionale Interuniversitario per le Telecomunicazioni (CNIT), Italy.\\}
\begin{document}
\begin{abstract}
Existing network stacks tackle performance and scalability aspects by relying on multiple receive queues. However, at software level, each queue is processed by a single thread, which prevents simultaneous work on the same queue and limits performance in terms of tail latency.
To overcome this limitation, we introduce COREC, the first software implementation of a concurrent non-blocking single-queue receive driver.  By sharing a single queue among multiple threads, workload distribution is improved, leading to a work-conserving policy for network stacks. On the technical side, instead of relying on traditional critical sections---which would sequentialize the operations by threads---COREC coordinates the threads that concurrently access the same receive queue in non-blocking manner via atomic machine instructions from the Read-Modify-Write (RMW) class. These instructions allow threads to access and update memory locations atomically, based on specific conditions, such as the matching of a target value selected by the thread. Also, they enable making any update globally visible in the memory hierarchy, bypassing interference on memory consistency caused by the CPU store buffers.
Extensive evaluation results demonstrate that the possible additional reordering, which our approach may occasionally cause, is non-critical and has minimal impact on performance, even in the worst-case scenario of a single large TCP flow, with performance impairments accounting to at most 2-3 percent. Conversely, substantial latency gains are achieved when handling UDP traffic, real-world traffic mix, and multiple shorter TCP flows.
\end{abstract}
\maketitle

\section{Introduction}
Modern online services are designed focusing on latency as one of the main performance indicators, not only in terms of average latency but also, and mainly, in terms of tail latency. In fact, this is especially challenging as the size and complexity of the system increase, since even rare hiccups can affect a significant portion of the workload \cite{dean2013tail}.


Latency variability in data centers has been widely studied \cite{dean2013tail,li2014tales}, and there are many possible sources of tail latency, like resource sharing, queuing, and power management. In more details, the literature shows that we have two sides of the same coin. On one side, we have to deal with latency variability while dealing with complex systems 
and cannot get entirely rid of it \cite{dean2013tail}. On the other one, it is crucial to build  architectures explicitly  designed to mitigate it as much as possible \cite{li2014tales, caladan, shenango, mcclurensdi22,userlevelthreding, qin2018arachne,perfiso,zygos, kaffes2019shinjuku, ix}.

In this context, we believe modern network stacks are not designed to deliver the best performance regarding tail latency. On the receive side, a network stack is made of different receive queues, where each queue permits software to exchange packets with the NIC and process them. However, in common software releases, one queue is processed by only one thread, and the threads cannot simultaneously work on the same queue. We can therefore model the network stack as a $N \times M/G/1$ queuing system, where $N$ threads manage a separate queue each. It is well known from basic queuing theory that an $M/G/N$ system would bring significant advantages in tail latency (see Section \ref{sec:motivation}); in fact, a single queue shared among the $N$ threads enables a global visibility of the workload to be processed by all the threads, therefore implying a work-conserving policy for network stacks.\\
We imagine two reasons why this approach has never been explored before:
\begin{enumerate}
    \item A shared queue implies the possibility of breaking per-flow consistency, with packets in the same flow ending up (simultaneously) in different threads. It also introduces the case of packet reordering in flow-based streams;
    \item Re-architecting the stack's queue policy requires the network driver's modification. Drivers are typically developed by the NIC vendor, and users typically treat them as a black-box component, knowing very little about how they work.
\end{enumerate}

From a pragmatic perspective, the software logic that queuing systems rely on admits a single execution flow at any time for managing the data-structure implementing the queue. In particular, even though the literature has proposed solutions \cite{metronomeconext} where multiple threads can process the packets incoming from a single queue, only one of these threads is enabled to carry out the actual operations at any time, thanks to a queue-locking mechanism.

We believe the aforementioned  motivations and the practical way to proceed with queue management---in particular, the reliance on locking and critical sections for the operations on a single queue---are now obsolete for the following reasons:
\begin{enumerate}
\item A new generation of concurrent algorithms, known as lockless algorithms, has garnered considerable attention due to their ability to handle shared data structures without relying on locking mechanisms, and thus enhance scalability and performance. However, while concurrent lock-free and wait-free approaches have been largely investigated by the Operating Systems community (see e.g., \cite{matveev2015read, marotta2021nbbs, natarajan2014fast, chatterjee2014efficient, harris2001pragmatic, DBLP:journals/tpds/IanniPQ19}), the networking community has given a significantly lower attention to this trend, to the extent that, to the best of our knowledge, they are currently not 
exploited at the level of any mainstream queue management driver.
    
\item Most of the data-center flows are restricted to a handful of packets \cite{homa, dagger}, therefore minimizing both the possibility and the effects of packet reordering when adopting concurrent (e.g. simultaneous) management of the queue by multiple-threads.

\item Recently, the networking community has shown some interest in network drivers \cite{emmerich2019user, pirelli2020simpler, emmerich2019case}. This has permitted the community to dive deeper on how a driver operates under the hood and on how
the actual software execution flow could further optimize it.
\end{enumerate}

We build on these three observations and present the first implementation (to the best of our knowledge) of a work-conserving, parallel network driver, where threads use no locking mechanism for managing the data structure, implementing a single queue. Therefore, our driver follows a scale-up policy for network stacks, which is fully orthogonal---and mixable with---the widely adopted scale-out policy.

This paper provides the following core  contributions:
\begin{itemize}
\item We present an algorithm fully supportable with any common ISA, like the one offered by {\sc x86} processors, where multiple concurrent threads can, at the same time, process different (sets of) packets incoming from the same queue. In this solution, thread coordination---for avoiding inconsistencies in the management of the queue data structure---purely occurs via the exploitation of Read-Modify-Write (RMW) machine instructions.

\item Our algorithm has been implemented on top of {\sc x86}/Linux machines and has been integrated within the DPDK packet processing framework \cite{dpdk}.

\item We present a comprehensive assessment of the capabilities that are permitted by our solution, compared to the classical literature scenario where every single queue is instead managed by an individual execution flow (a single thread) at any time. Our approach has positive benefits in terms of both mean and tail latency, in particular when used on an L3 forwarder. We then show that the possibility of reordering is minimal for typical flow-size packets; thus, it has a minimal impact on very long TCP flows, while it enables significant latency benefits for medium and short flows.
\end{itemize}

The remainder of this article is structured as follows. Section \ref{sec:drivers} provides a background on network drivers. In Section \ref{sec:nonblockingdriver}, our concurrent non-blocking single-queue receive driver---which we named COREC---is presented, discussing both  theoretical and practical aspects. In Section \ref{sec:eval}, a comprehensive evaluation of COREC is presented. Finally, related work is discussed in Section \ref{sec:rw}.

\section{Network drivers} \label{sec:drivers}
\subsection{Baseline concepts}

\begin{figure}[]
    \centering
    \includegraphics[width=0.45\textwidth]{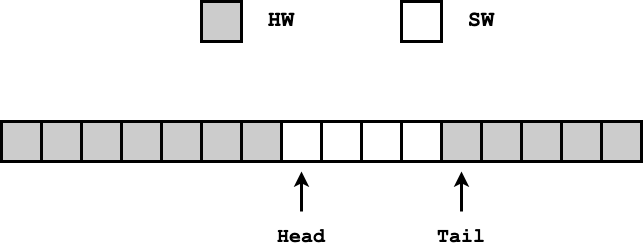}
    \caption{A typical ring buffer}
    \label{fig:ringbuffer}
\end{figure}

A network driver is no more than a piece of software built to manage packets incoming/outgoing from/to the NIC.
One or more circular buffers (also called Rx/Tx queues) are shared between the NIC and the CPUs. Furthermore, incoming traffic can be split into multiple Rx queues through filters or a hashing algorithm, while multiple Tx queues are usually merged on the NIC. These queues are composed of descriptors, each containing some metadata about the packet and a pointer to a memory area where the packet is located. An Rx queue routine is roughly like this (the Tx one is specular): the NIC controls the area between the tail and the head of the queue \cite{inteldatasheet} (see the grey boxes in Figure \ref{fig:ringbuffer}) and moves the head forward for every descriptor it fills. Complementarily, the software controls the rest of the queue, swapping the NIC-populated descriptors with empty ones and moving the tail. The software can understand whether a descriptor is populated or not through the DD bit, which the NIC sets once it fills the descriptor.

\subsection{Rx side flow}

For the reader's convenience, we now show a simplified routine of the receive side of a network driver  in Listing \ref{lst:recv}. 
First, the driver retrieves the relevant pieces of information, namely the buffer that is shared  with the NIC (line 6) and the descriptor from where it left off at the last iteration (line 7). The driver checks how many of the following descriptors were populated by hardware by reading their DD bit (line 10) up to a fixed batch value (usually 32). Each of these descriptors is moved to a user-space buffer (line 15) and replaced with a new one taken from a pre-allocated memory pool (line 17). Software can now update the tail, moving it by the number of packets it has retrieved (line 24).
\begin{lstlisting}[style=example, caption={A standard receive function of a network driver}, label={lst:recv}]
#define wrap_ring(index) (uint16_t) (index % RING_SIZE)

uint32_t ixgbe_rx_batch(struct device* dev, uint16_t queue_id, struct pkt_buf* bufs[]) {
  //Get the queue struct for device dev and queue queue_id 
  struct ixgbe_rx_queue* queue = get_queue(dev, queue_id);
  struct pkt_buf* buffer = queue->buffer;
  uint16_t rx_index = queue->rx_index; // descriptor index we checked in the last run of this function
  uint16_t last_rx_index = queue->rx_index;
  for (int i = 0; i < BATCH_SIZE; i++) {
    if (!(buffer[rx_index] && DD_BIT))
    //Descriptor has not been filled by the NIC yet, exit the loop
      break;
    else {
      //Copy the descriptor in the user space buffer
      bufs[i] = buffer[rx_index];
      //Replace the descriptor with a new one from the mempool
      buffer[rx_index] = mempool_desc_get();
      last_rx_index = rx_index;
      rx_index = wrap_ring(rx_index + 1);
      
    }
  }
  //Free the processed descriptors back to the NIC
  set_register(TAIL, dev, queue_id, last_rx_index)
}
\end{lstlisting}
\section{The concurrent non-blocking receive driver} \label{sec:nonblockingdriver}

\subsection{Core concepts}\label{sec:idea}
This paper's main idea is to design a non-blocking algorithm that can enable a multithreaded, simultaneous processing of a {\em same} Rx queue---hence avoiding any locking of the queue.
From the previous discussion, it is clear that the receive flow of a network driver is anything but tailored to a concurrent execution model: in fact, the NIC-shared data structures lack the support for simultaneous operations in order to access them in parallel without causing inconsistencies (and therefore, malfunctions) in the buffer state.
This causes the whole Rx function to be a critical section.

This article proposes a different approach based on the core concepts that drive the development of modern concurrent algorithms \cite{matveev2015read, marotta2021nbbs, natarajan2014fast, chatterjee2014efficient, harris2001pragmatic} arguably overlooked by the networking community.
In 
 these  solutions, the notion of atomicity (hence  correctness) of the operations by a thread is  no longer linked to the concept of critical section. Rather, threads are coordinated by the reliance on atomic machine instructions belonging to the Read-Modify-Write (RMW) class. 
 These instructions can  access a memory location and update it---for example if the original value matches a target value selected by the thread. 
 
 
 Some of these instructions can fail in the update operation---e.g. if the memory location does not (or no longer) match the target value that has been selected.
 Hence, the failure makes the thread know that the shared data structure has changed its state---e.g. because of operations occurring by a concurrent thread. 
Threads can therefore fail/win a race in constant time, 
and in case of a win, the thread has earned the right to perform a specific operation, which is immediately visible to the other threads (i.e. they fail) so that race conditions are avoided\footnote{Full visibility of the operations by RMW instructions is supported by the setup of the atomicity of their execution with respect to the flush of store buffers of the CPUs towards the 
cache/RAM architecture.}.
In case of a fail, a thread has not modified the shared state and hasn't caused any delay or inconsistency for the concurrent workers.
As a result of this design, the threads are totally decoupled, enabling total independence (they do not block each other) and resilience against slowdowns (e.g., de-scheduling, cache misses, and interrupts). 

\begin{figure}[]
    \centering
	\includegraphics[width=0.45\textwidth]{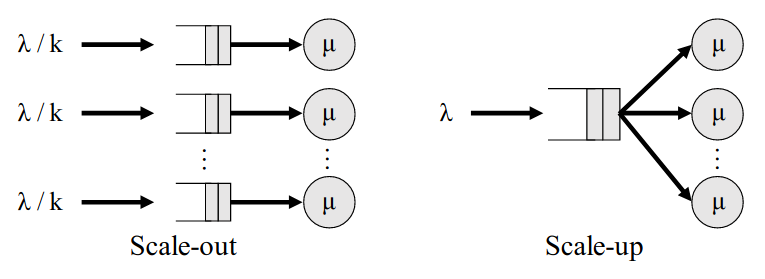}
	\caption{Scale-out (NxM/M/1) vs. scale-up (M/M/N) policy}
	\label{fig:scale}
\end{figure}

The direct consequence is the possibility of a scale-up mechanism in current network functions and end-hosts, opposite to the current scale-out policy (Figure \ref{fig:scale}). 

We note that this way of designing concurrent algorithms is de-facto an advancement in terms of how these RMW machine instructions can be exploited. In fact, they have been largely 
used for long time to implement locking---like spinlocks that can be atomically taken by a single winner at any time. The literature on non-blocking algorithms migrates the usage of RMW instructions at a different level, embedding it into the actual algorithm that accesses and manipulates the shared data structure. This is the path also followed by our network driver solution.

\subsection{Theoretical aspects}\label{sec:motivation}
The concept of using multiple threads to process a single shared queue is also well-grounded when considering queuing theory. In fact, this technique provides a global perspective of the workload distribution among threads, which leads to a work-conserving strategy that can handle temporal traffic imbalances and head-of-the-line blocking. 

To evaluate the effectiveness of this approach from a theoretical perspective, we conducted several simulations using the Matlab Simevents package, which incorporated Markovian arrival rates and both Markovian and Deterministic service times, while varying the number of servers (4 and 8). 

\begin{figure*}[h!]
\centering
\captionsetup{justification=centering}
\begin{minipage}{0.24\textwidth}
  \centering
\includegraphics[width=\textwidth]{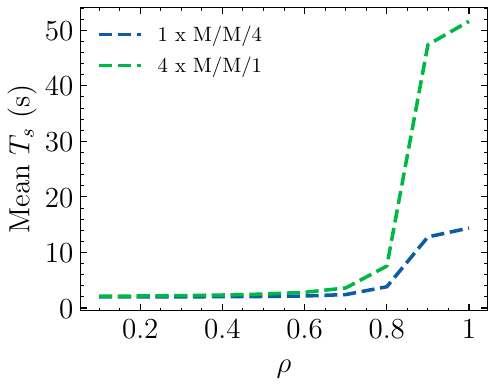}
\subcaption{Mean latency -\\ 4 cores}
\end{minipage}%
\begin{minipage}{0.24\textwidth}
  \centering
\includegraphics[width=\textwidth]{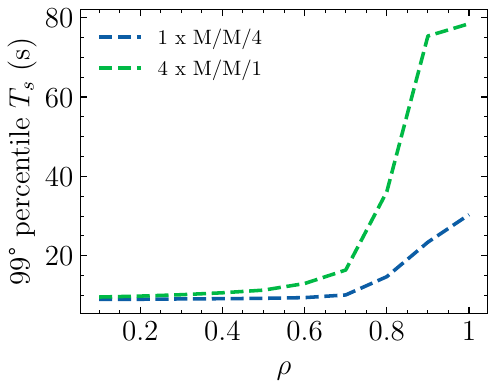}
\subcaption{99p latency -\\ 4 cores}
\end{minipage}%
\begin{minipage}{0.24\textwidth}
  \centering
\includegraphics[width=\textwidth]{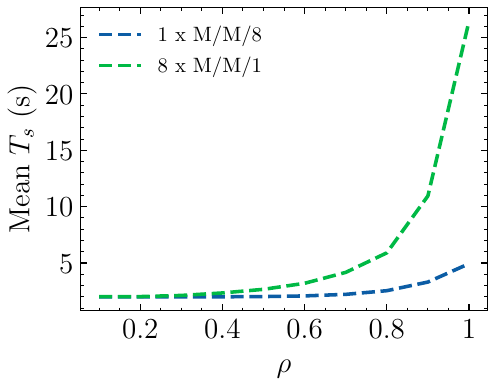}
\subcaption{Mean latency -\\ 8 cores}
\end{minipage}
\begin{minipage}{0.24\textwidth}
  \centering
\includegraphics[width=\textwidth]{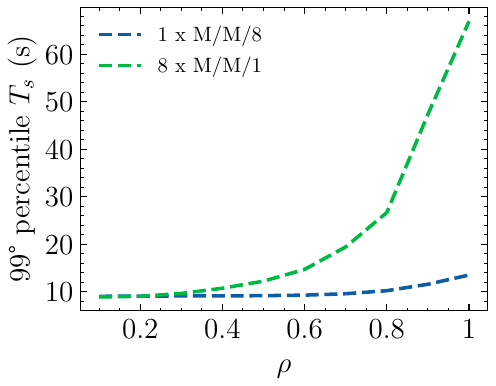}
\subcaption{99p latency -\\ 8 cores}
\end{minipage}
\caption{Mean and 99p latency simulation results - Markovian service time} \label{fig:motivation_M}
\end{figure*}

\begin{figure*}[h!]
\centering
\captionsetup{justification=centering}
\begin{minipage}{0.24\textwidth}
  \centering
\includegraphics[width=\textwidth]{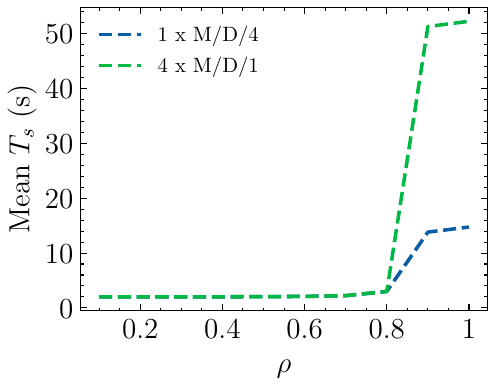}
\subcaption{Mean latency -\\ 4 cores}
\end{minipage}%
\begin{minipage}{0.24\textwidth}
  \centering
\includegraphics[width=\textwidth]{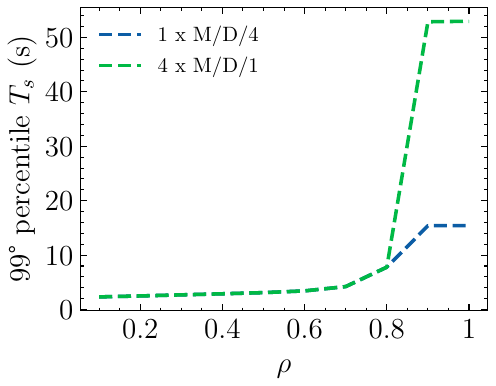}
\subcaption{99p latency -\\ 4 cores}
\end{minipage}%
\begin{minipage}{0.24\textwidth}
  \centering
\includegraphics[width=\textwidth]{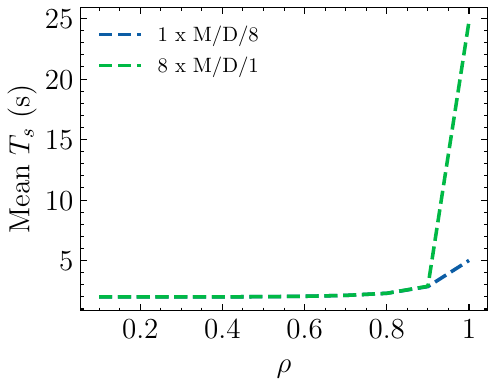}
\subcaption{Mean latency -\\ 8 cores}
\end{minipage}
\begin{minipage}{0.24\textwidth}
  \centering
\includegraphics[width=\textwidth]{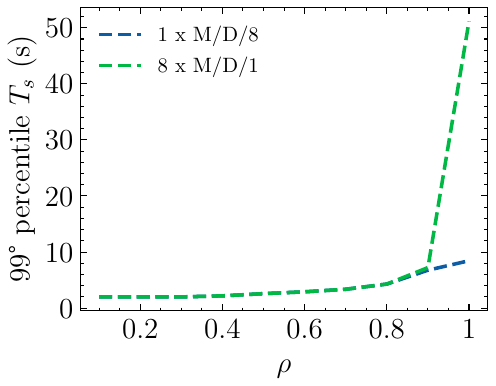}
\subcaption{99p latency -\\ 8 cores}
\end{minipage}
\caption{Mean and 99p latency simulation results - Deterministic service time} \label{fig:motivation_D}
\end{figure*}

The results, presented in Figure \ref{fig:motivation_M}, display both mean and 99p latency. Figure \ref{fig:motivation_M} shows results for a Markovian service time. 
For each plot, the blue line shows our approach (\textbf{scale-up, or M/M/N}), while the green one shows the current state-of-the-art (\textbf{scale-out, or NxM/M/1}).
{\bf Our findings clearly demonstrate that using multiple threads aligns with queuing theory and significantly improves both mean and tail latency.}
We now repeat the same test with Deterministic service times, a scenario highly unlikely to happen in modern computing systems because of the presence of multiple sources of variability. Still, this utopian scenario represents the case with fewer benefits for the proposed approach. The results are shown in Figure \ref{fig:motivation_D}; it is interesting to underline that our approach still brings benefits at a very high load.

In a real scenario, the core point stands in how to build the multithreaded concurrent queuing system in an effective manner---for example via the well suited exploitation of machine instructions in the RMW class. Challenges and constraints related to this aspects are discussed in the next section.

\subsection{Challenges and constraints}\label{sec:constraints}
Let us go back to Listing \ref{lst:recv} to understand the portions of code where race conditions among concurrent threads can occur:
\begin{enumerate}
    \item in lines 15-17, a concurrent copy and replace of the same descriptor can happen from more threads at the same time, causing inconsistencies in the shared buffer. 
    \item at line 24, the TAIL write is dependent on the timing of the thread's operations. 
\end{enumerate}
There is also a fundamental question to be investigated: for how the problem has been presented until now, the reader might think that concurrent algorithms for accessing a shared ring buffer have already been implemented in software and, therefore, may not see any significant contribution. The main difference with respect to the existing algorithms is that in the latter case, both producers and consumers are written by the user in customized software, while in our case, we can only write the consumer's portion of code without having any possibility to modify the producer's behavior, which is the NIC. Consequentially, there is also the need to make our algorithm compatible with what the NIC is expecting, namely a single execution flow coherently processing the receive queue.
This is both a requirement for running our driver with unmodified NICs and a limitation on how the threads can behave since the NIC must not notice that multiple threads are simultaneously processing the same queue.

\subsection{The algorithm}
\subsubsection{Handling thread-level parallelism}
Let's see how we can overcome the above-mentioned conflicts through Listing \ref{lst:mt1}: 
\begin{enumerate}
    \item \textbf{Split the work to be done:} the set of the available (in the sense of containing a packet) descriptors must be \textit{partitioned} disjointedly among the threads, so that they don't overlap.
    This is achieved through the following operations: first, a scan of the Rx queue is done by reading the DD bit (lines 12-19) in order to understand how many descriptors have been filled by the NIC. At this point, the thread tries to obtain that specific batch of descriptors through an atomic Compare-And-Swap (CAS) machine instruction of the RMW class (line 21). In case of a win, the \texttt{queue->rx\_index} global variable has been instantaneously updated, and therefore no other thread can obtain a conflicting set of descriptors\footnote{Each conflicting thread has two scenarios: either it sees the new value of \texttt{queue->rx\_index} when getting a copy at line 8 or if they still have the old one, they will fail the race for modifying it at line 21.}.
    In lines 23-30, the thread can copy the descriptors to its own buffer and replace them with new, empty ones. This is the actual portion of code we can speed up in this execution model.
    \item \textbf{Synchronize on who should update the TAIL register:} we avoid concurrent TAIL writes through a trylock (lines 35 and 42). We underline that even if the \texttt{trylock()} call fails there are no negative consequences for the thread in terms of waiting or delay.
\end{enumerate}
\subsubsection{Handling transparency to the NIC}
 The above-mentioned NIC compatibility problem calls for a \textit{transparency} mechanism, which is a way for making 
 threads simultaneously  process the same queue while giving the illusion to the NIC that it is interfacing with only one thread. More specifically, with \textit{transparency} we mean
that the set of descriptors one can re-assign to the NIC \textit{must} be contiguous; in this way, the NIC will see the re-assigned descriptors as a unique batch released by a single thread. As an example: say thread A has granted descriptor 1 and then thread B has granted descriptor 2, but thread A is being slowed down for some reason while B has ended its work. In this case, one can't just write 2 to the TAIL register as this would also mean freeing descriptor 1, which is not done yet. At this point, thread B should wait for thread A for an unknown amount of time, and we have already stated in Section \ref{sec:idea} that we \textit{don't} want this. So the only thing B can do to exit the Rx function and process the packet it has received in the meantime is to write to some shared data structure that descriptor 2 can be freed, otherwise, this information would be lost. When thread A eventually ends its routine, it will first write to the same data structure that descriptor 1 is done. Then if there is a contiguous set of descriptors (starting from the current tail onwards) that can be re-assigned to the NIC, it will understand this from reading this shared data structure and will eventually move the TAIL. We could still have multiple threads trying to read this shared data structure and concurrently trying to update the TAIL. Still, the idea is that a thread that sees a continuous batch of descriptors (either by "filling a gap" as thread A did in the previous example or by creating a brand new one) can give them back to the NIC by updating the TAIL.
\subsubsection{The practice}
\begin{lstlisting}[style=example, caption={A simplified receive function of a parallel network driver}, label={lst:mt1}]
#define wrap_ring(index) (uint16_t) (index % RING_SIZE)
uint16_t lock = 0;

uint32_t ixgbe_rx_batch(struct device* dev, uint16_t queue_id, struct pkt_buf* bufs[]) {
  //Get the queue struct for device dev and queue queue_id 
  struct ixgbe_rx_queue* queue = get_queue(dev, queue_id);
  struct pkt_buf* buffer = queue->buffer;
  uint16_t rx_index = __atomic_load(queue->rx_index); // descriptor index we checked in the last run of this function
  //Local copy of the rx_index counter
  uint16_t rx_index_local = rx_index;
  uint16_t last_rx_index, i;
  for (i = 0; i < BATCH_SIZE; i++) {
    if (!(buffer[rx_index_local] && DD_BIT))
    //Descriptor has not been filled by the NIC yet, exit the loop
      break;
    else 
      //Move on to the next descriptor
      rx_index_local = wrap_ring(rx_index_local + 1);
  }
  //try to win the race for the batch of descriptors [rx_index... rx_index_local]
  if (__compare_and_swap(&queue->rx_index, rx_index, rx_index_local)) {
    //race is won, we can copy the descriptors
    rx_index_local = rx_index;
    new_bufs = mempool_desc_bulk_alloc();
    for (uint16_t j = 0; j < i; j++) {
      bufs[j] = buffer[rx_index_local];
      //Replace the descriptor with a new one from the mempool
      buffer[rx_index_local] = new_bufs[j];
      last_rx_index = rx_index_local;
      rx_index_local = wrap_ring(rx_index_local+1);
    }
    //write that the [rx_index... rx_index_local] is successfully copied
    write_batch_is_done(rx_index, rx_index_local);
  }
  if (trylock(&lock)) {
    //Get how many contiguous descriptors there are to be freed, starting from the TAIL
    uint16_t descs_to_free = read_batch_done(queue->tail);
    //Set the descriptors' bits back to 0
    write_batch_to_zero(queue->tail, wrap_ring(queue->tail + descs_to_free))
    //Free the processed descriptors back to the NIC
    set_register(TAIL, dev, queue_id, wrap_ring(queue->tail + descs_to_free));
    release_lock(&lock);
  }
}
\end{lstlisting}
In the deployment of our approach, we have found many practical situations that must be taken into account; these are presented here:
\begin{enumerate}
    \item \textbf{Global transaction ID:} we need some unique ID that tells us where the process of assigning descriptors to cores has arrived and that we can update through the CAS operation at line 21 in Listing \ref{lst:mt1}; how do we choose it?\\Unfortunately, the naive choice of using the Rx queue descriptor index \texttt{queue -> rx\_index} at line 8 cannot be done. The reason behind this is that the ID, since it ranges from 0 to \texttt{RING\_SIZE-1}, is susceptible to the ABA problem; thread A may read index 1023, be descheduled and after some other thread has done a complete round of processing the queue (so the ID is now 1023 again), thread A may wake up again and successfully do a CAS operation, even if it saw an ancient state of the queue! The only solution is, therefore, to use a constantly increasing ID in order to make impossible this periodic wrapping of the index (e.g., using an unsigned 32-bit integer). The assumption here is that the queue size is always a power of 2 to map the ID to the queue positions correctly, but this already happens in network drivers, so we're not limiting the possible Rx queue size in any way. When overflow occurs, the variable will start again from 0, and this does not cause any inconvenience. For mapping the ID to the descriptor offset in the queue, we just need to divide it by the queue size and get the rest of the division. The result of the division tells us another piece of information, namely the \textit{epoch} in which the queue is (See Table \ref{tab:epoch}). The epoch means how many times the system has done a complete round in processing that queue, so from 0 to SIZE-1. The critical point here is that choosing an ever-growing ID allows us to distinguish between the different epochs the queue may be into, avoiding the previous problem.
    \item \textbf{How to store in a shared data structure whose descriptor has been processed (by any thread) and is ready to be assigned to the NIC?:} We choose to use a bitmask with one bit per descriptor called \texttt{READ\_DONE}. This permits us to do the following thing: when all descriptors belonging to a certain iteration have been processed, the thread knows which bits it has to write, and this likely translates into an atomic write to a single variable: in line 33 at Listing \ref{lst:mt1}, the thread writes the batch starting from \texttt{rx\_index} to \texttt{rx\_index\_local}. Bits need not only to be set to 1 at the end of the processing, rather they also need to be set back to 0 (line 39) when a thread grants responsibility for freeing certain descriptors to the NIC (line 41); otherwise, this would cause conflicting views.\\
\end{enumerate}
\begin{table}[]
\centering
\begin{tabular}{c|c|c}
ID   & Descriptor Index & Epoch               \\ \hline
0    & 0                & \multirow{5}{*}{0}  \\
1    & 1                &                     \\
2    & 2                &                     \\
...  & ...              &                     \\
1023 & 1023             &                     \\ \hline
1024 & 0                & \multirow{5}{*}{1}  \\
1025 & 1                &                     \\
1026 & 2                &                     \\
...  & ...              &                     \\
2047 & 1023             &                     \\ \hline
2048 & 0                & \multirow{5}{*}{2}  \\
2049 & 1                &                     \\
2050 & 2                &                     \\
...  & ...              &                     \\
3071 & 1023             &                     \\ \hline
\end{tabular}
\caption{Table with the global transaction ID (left), the referred descriptor index (centre) and the consequent epoch (right)} \label{tab:epoch}
\end{table}

\subsubsection{Corner cases} \label{sec:corner}
The main corner case that may cause the system to stall is preventing the NIC from loading incoming packets to the shared buffer, i.e., the buffer is full. More specifically, say thread A has granted possession of descriptor 2 and gets descheduled for an indefinite period of time. In this situation, other threads can process a full round of descriptors (from 3 to 1).  Then they will always find the queue full of new descriptors ready to be re-assigned to the NIC but unable to be returned to the NIC because they lack descriptor 2 in order to form a contiguous batch of descriptors starting from the TAIL. In the meantime, the NIC sees the buffer as full since no thread has had the possibility to move the TAIL. Thus, threads will have to wait for A to resume its execution and mark descriptor 2 as ready to be assigned back to the NIC in the \texttt{READ\_DONE} shared variable. This is not a limitation strictly caused by our proposed approach but instead by the way in which the NIC-to-CPU communication is designed (see Section \ref{sec:constraints}). We underline that, with respect to the opposed scale-out policy, this approach still permits to perform a full round of operations on the shared buffer, while in the state-of-the-art scale-out policy if one thread is delayed then the whole receive queue(s) assigned to it cannot be processed in any way.
\subsection{Implementation}
The new network driver routine has been implemented in DPDK v21.11\footnote{We will make the code available for the final version of the paper}. Our contribution is not restricted to DPDK but could be extended to other frameworks like the Linux kernel, XDP, or RDMA completion queues. We chose DPDK since it gave us the possibility of writing C code in user space, also enabling easy deploying and debugging activities.\\
We focused on the \texttt{ixgbe}, \texttt{i40e} and \texttt{ice} Intel drivers. We chose such drivers since the \texttt{ixgbe} driver is well-documented \cite{inteldatasheet} and well-explained by Emmerich et al. in \cite{emmerich2019user} through their simplified \texttt{ixy} driver; \texttt{i40e} and \texttt{ice} are also pretty similar to \texttt{ixgbe} in terms of how the receive function works. Other vendors tend not to show their drivers' routines and specifications publicly through datasheets: we encourage them to make this information public in order to increase researchers' interest and knowledge in network drivers.\\
DPDK drivers usually exploit vectorized ASM instructions in the Rx routine in order to optimize performance further, but unfortunately, these are not documented; for this reason, we disabled the vectorized receive function versions and focused only on the standard ones.\\
Regarding the actual code writing, RMW-based coordination of the threads is achieved through the \texttt{\_atomic} \cite{atomic} and \texttt{\_sync} \cite{sync} \texttt{gcc} built-in primitives. \texttt{\_atomic} functions avoid reordering from out-of-order execution, while the \texttt{\_sync\_bool\_compare\_and\_swap} function performs as a test-and-set primitive: it allows us to atomically control if a specific memory location matches a particular value and if the two match, to update the location to a new value.

\section{Evaluation}\label{sec:eval}
In this section, we present the evaluation tests for COREC, our concurrent non-blocking single queue receive driver. COREC is compared against the standard DPDK scale-out policy (v21.11). 
Section \ref{sec:scalability} focuses on how multiple threads scale when bound to the same queue, Section \ref{sec:latency} shows the results for mean and 99p latency while Section \ref{sec:reordering} quantifies the packet reordering percentage for different traffic sizes. Tests are executed on a server equipped with Intel Silver Xeon 4110 CPU clocked at 2.1GHz, Intel XL710 40Gbps NICs and X520 10Gbps NICs. CPU cores are isolated, and power limitators, like C-states and P-states, are disabled. The server is running Linux Kernel v5.13. The sender uses traffic generators like MoonGen \cite{moongen} or Trex, depending on the test scenario. MoonGen fails to saturate 40Gbps NICs with 64B packets, but it still provides essential features like hardware rate control with Intel X520 NICs (which is not available with Intel XL710) in order to properly quantify the reordering rate. This is why, depending on the scenario, we vary both the traffic generator and the NIC used. In all of the scale-out cases, the traffic flow distribution is equal among cores.\\
The tested applications are examples included the DPDK framework, namely:
\begin{itemize}
    \item \texttt{l3fwd} \cite{l3fwd}, which acts as a Layer-3 longest-prefix-matching forwarder. The application retrieves packets from the NIC in batch, executes a routing table lookup for each packet, and forwards it through the relevant NIC. This NF permits us to retrieve the per-packet latency easily.
    \item \texttt{ipsec-gw} gateway \cite{ipsecgw}, which is a more expensive task since it performs more complex operations; after retrieving a batch of packets, it controls for every packet which rules to apply based on its flow (forward, drop, encryption/decryption), performs the operations and sends the packet if required. In our setup, one third of the packets are forwarded, one third are dropped and the other third are encrypted in software without NIC HW acceleration and then forwarded. We expressly avoid the HW acceleration features to emulate a more complex NF in terms of CPU cycles, and therefore see how our driver scales in this case.
\end{itemize}
\subsection{Scalability tests}\label{sec:scalability}
A first metric worth being measured is how our driver scales in throughput when we add more threads to the same queue, as well as how our driver . We show the results in Table \ref{tab:scalabilityl3fwd} for \texttt{l3fwd} and Table \ref{tab:scalabilityipsec} for \texttt{ipsec-gw}. We show the throughput in Mpps for 64B traffic when executing the tasks on a NIC-local NUMA node and a remote one, as well as the performance improvement in percentage compared to the state of the art. In the case of Table \ref{tab:scalabilityl3fwd}, the NIC maximum throughput is around 37 Mpps as stated in the Intel XL710 datasheet \cite{intelxl710}, so a hardware limitation biases this value. A first observation is that our algorithm, also in the 1:1 thread-to-queue comparison, provides some benefits in throughput: the reason behind this is the use of a bulk allocation mechanism from the memory pool (Line 24 in Listing \ref{lst:mt1}). A second observation is that our algorithm shows better scalability improvements in the remote NUMA scenario; this is a direct consequence of the increased memory access time latencies.

\begin{table}[]
\tabcolsep=0.11cm
\footnotesize
\begin{tabular}{|l|cc|cc|}
\hline
mode & \multicolumn{2}{|c|}{same NUMA} & \multicolumn{2}{c|}{different NUMA} \\ \hline
 & Tput (Mpps) & \% & Tput (Mpps) & \% \\ \hline
\textbf{DPDK}              & 16.43 & 100         & 11.54 & 100 \\ \hline
\textbf{COREC 1 core}  & 17.68 & 107.61 & 11.87 & 102.86             \\ \hline
\textbf{COREC 2 cores} & 26.4  & 160.68 & 19.62 & 170.02              \\ \hline
\textbf{COREC 3 cores} & 35.35 & 215.16 & 28.09 & 243.41             \\ \hline
\textbf{COREC 4 cores} & 37.66 & 229.21 & 33.69 & 291.94             \\ \hline
\end{tabular}
\caption{Scalability executing L3FWD task}
\label{tab:scalabilityl3fwd}
\end{table}

\begin{table}[]
\tabcolsep=0.11cm
\footnotesize
\begin{tabular}{|l|cc|cc|}
\hline
mode & \multicolumn{2}{|c|}{same NUMA} & \multicolumn{2}{c|}{different NUMA} \\ \hline
 & Tput (Mpps) & \% & Tput (Mpps) & \% \\ \hline
\textbf{DPDK}              & 5.04 & 100         & 2.74 & 100 \\ \hline
\textbf{COREC 1 core}  & 5.07 & 100.6 & 2.81 & 102.55             \\ \hline
\textbf{COREC 2 cores} & 8.18  & 162.3 & 4.94 & 180.29              \\ \hline
\textbf{COREC 3 cores} & 10.9 & 216.27 & 6.92 & 252.55             \\ \hline
\textbf{COREC 4 cores} & 15.3 & 303.57 & 8.92 & 325.52             \\ \hline
\end{tabular}
\caption{Scalability executing IPSec task}
\label{tab:scalabilityipsec}
\end{table}
\subsection{Latency}\label{sec:latency}
\begin{figure*}
\begin{minipage}{0.45\linewidth}
  \centering
	\includegraphics[width=\textwidth]{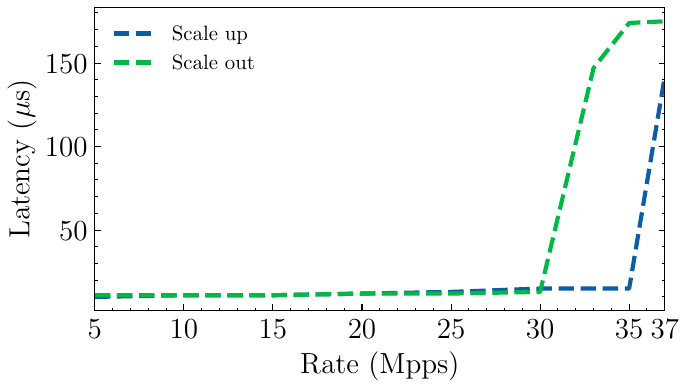}
	\subcaption{4 cores}
	\label{fig:meanlatencyl3fwd4cores}
 \end{minipage}
\hspace{0.2cm}
\begin{minipage}{0.45\linewidth}
\centering
	\includegraphics[width=\textwidth]{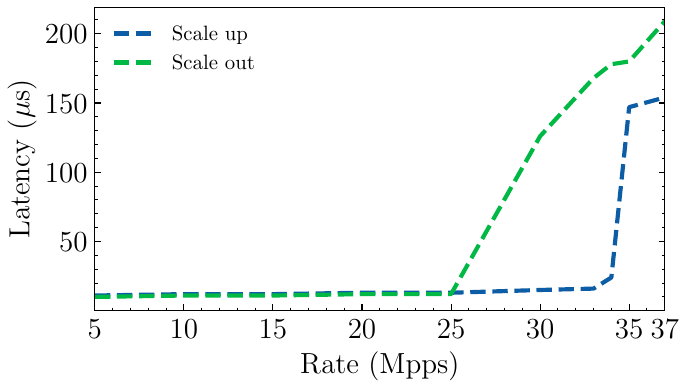}
	\subcaption{8 cores}
	\label{fig:meanlatencyl3fwd8cores}
\end{minipage}
\caption{L3FWD mean latency} \label{fig:latencymean}
\end{figure*}

We now focus on presenting the end-to-end latency benefits of our approach, motivated by the simulation results in Section \ref{sec:motivation}. Our scale-up approach and the state of the art (scale-out) are compared with different numbers of cores running the \texttt{l3fwd} task.
Figure \ref{fig:meanlatencyl3fwd4cores} and \ref{fig:meanlatencyl3fwd8cores} show the mean latency with 4 and 8 cores, respectively, and a variable load, from 0 to the maximum rate sustainable (37Mpps). We can see a similarity between the theoretical plots and the experimental results, as our approach maintains a flat mean latency until the system reaches the saturation point at 37Mpps.

\begin{figure*}
\begin{minipage}{0.45\linewidth}
  \centering
	\includegraphics[width=\textwidth]{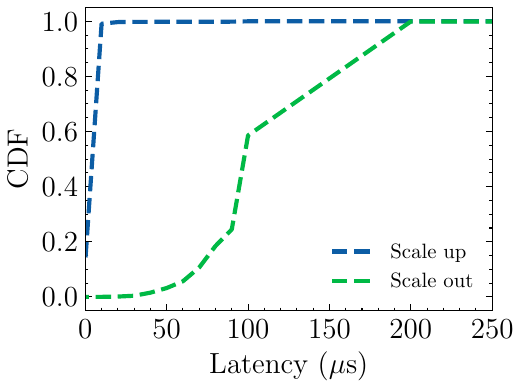}
	\subcaption{4 cores - 35Mpps}
	\label{fig:99platencyl3fwd4cores}
 \end{minipage}
\hspace{0.2cm}
\begin{minipage}{0.45\linewidth}
\centering
	\includegraphics[width=\textwidth]{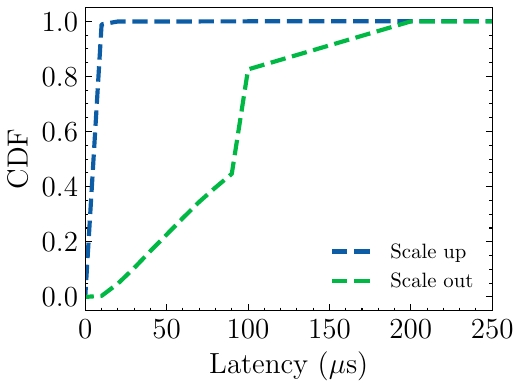}
	\subcaption{8 cores - 30Mpps}
	\label{fig:99platencyl3fwd8cores}
\end{minipage}
\caption{L3FWD latencies CDF distribution} \label{fig:latency99p}
\end{figure*}

We now focus on rates where our scale-up policy achieves better mean performance than the classical scale-out mechanism. Figure \ref{fig:99platencyl3fwd4cores} shows the CDF latency distributions at 35Mpps with 4 cores, while Figure \ref{fig:99platencyl3fwd8cores} shows the same lines with 8 cores and a 30Mpps traffic. The two figures clearly show that our approach brings benefits not only in terms of mean latency but, most of all, latency predictability.

\subsection{Reordering}\label{sec:reordering}
Tests are executed first with UDP traffic in order to retrieve the reordering metrics without the rate-sending limitations typical of TCP flows. Successive tests are executed with one/many TCP flow(s) in order to quantify the implications on real-life streams of data. We underline that COREC can be dynamically turned on or off (e.g. turn off can take place online by simply pausing the execution of the receiver threads, except one), so reordering might be a useful metric in order to decide whether to use it or not. 
\subsubsection{UDP}
\begin{figure*}
\begin{minipage}{0.45\linewidth}
  \centering
	\includegraphics[width=\textwidth]{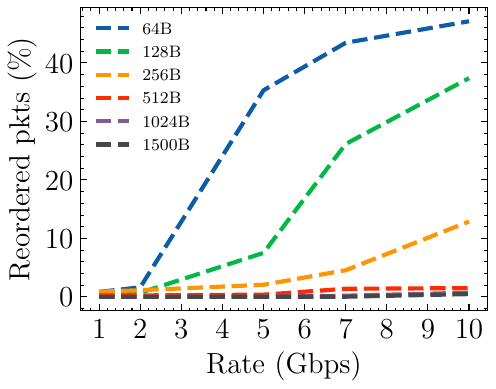}
	\subcaption{4 cores}
	\label{fig:udpreorder4cores}
 \end{minipage}
\hspace{0.2cm}
\begin{minipage}{0.45\linewidth}
\centering
	\includegraphics[width=\textwidth]{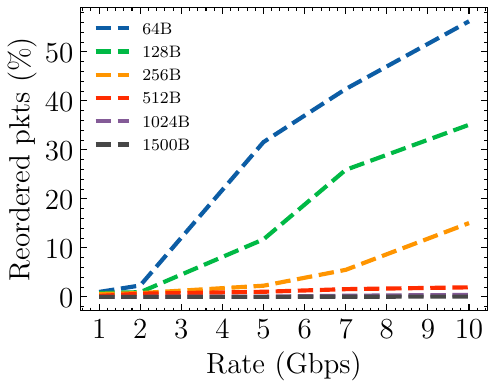}
	\subcaption{8 cores}
	\label{fig:udpreorder8cores}
\end{minipage}
\caption{Percentage of reordered packets for UDP traffic of different sizes} \label{fig:udpreorder}
\end{figure*}

Tests are performed following the metrics shown in the ``Packets Reordering Metrics'' RFC 4737 \cite{rfc4737}, namely the percentage of reordered packets. On a 10Gbps link, our test focuses on sending 100k sequenced packets, making them reordered by the COREC driver through a L3 forwarder, and checking the order of the arrived packets at the receiver side.
Tests are performed with different packet rates and sizes; results are shown in Figure \ref{fig:udpreorder4cores} with 4 cores pinned to the same Rx queue and in Figure \ref{fig:udpreorder8cores} with 8 cores.
It can be clearly seen that high levels of packet reordering are achieved only in presence of both high traffic rates and minimal packet sizes. In fact, as the packet sizes increase, the reordering percentage rapidly drops and becomes insignificant for typical flow sizes like TCP ones.

\noindent\textbf{Real-world traces:} We now repeat the same test with traces coming from real-world scenarios. These tests are particularly interesting because of the realistic characteristics of these traces, namely the different packet sizes and interarrival time between them. We used different daily traces from the MAWI (Measurement and Analysis on the WIDE Internet) Working Group available in \cite{mawitraces}: results are shown in Table \ref{tab:reorderrealworld}. It is clear that reordering has a minimal impact on real-life scenarios (all tests show that reordering packets are below 1\%). In this case, we also show the maximum distance between packets for each test.

\begin{table}[]
\tabcolsep=0.11cm
\footnotesize
\begin{tabular}{|c|ccc|ccc|}
\hline
Trace & \multicolumn{3}{|c|}{Pkts reordered (\%)} &\multicolumn{3}{c|}{Max distance}  \\ \hline
 & 2 cores & 4 cores & 8 cores & 2 cores & 4 cores & 8 cores\\ \hline
20210322 & 0.71925 & 0.93945 & 0.9981 & 31 & 34 & 34\\
20210323 & 0.28958 & 0.35312 & 0.46666 & 9 & 9 & 22\\
20210324 & 0.38129 & 0.55075 & 0.56831 & 8 & 43 & 45\\
\hline
\end{tabular}
\caption{Reordering results for MAWI traces}
\label{tab:reorderrealworld}
\end{table}
\subsubsection{TCP}
\begin{figure*}[]
\centering
\captionsetup{justification=centering}
\begin{minipage}{0.4\textwidth}
\includegraphics[width=\textwidth]{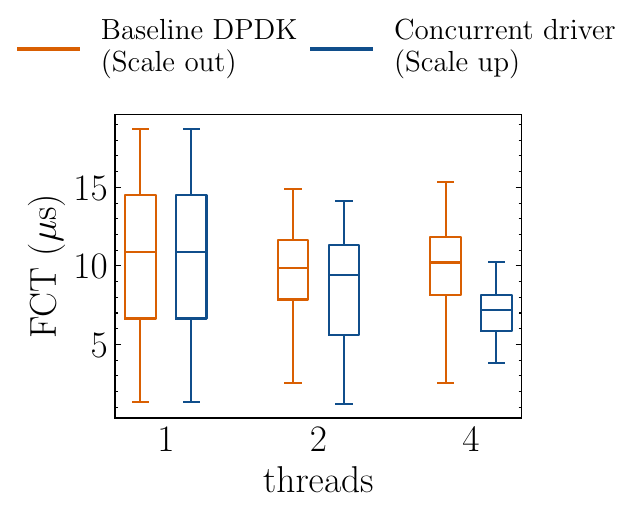}
\subcaption{64 flows}\label{fig:tcp_100k_64}
\end{minipage}
\begin{minipage}{0.4\textwidth}
\includegraphics[width=\textwidth]{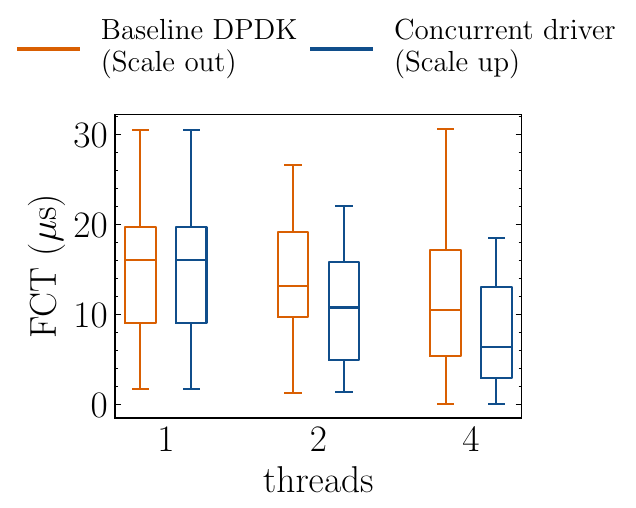}
\subcaption{128 flows}\label{fig:tcp_100k_128}
\end{minipage}
\caption{Flow Completion Time with 100KB payload}
\end{figure*}

\begin{figure*}[]
\centering
\begin{minipage}{0.4\textwidth}
\includegraphics[width=\textwidth]{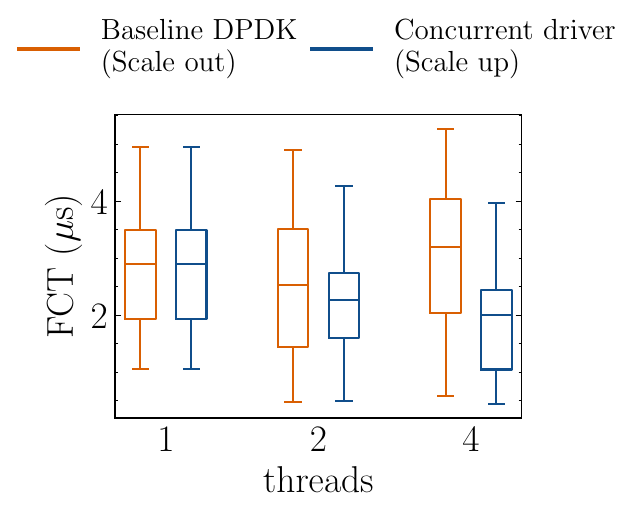}
\subcaption{64 flows}\label{fig:tcp_10k_64}
\end{minipage}
\begin{minipage}{0.4\textwidth}
\includegraphics[width=\textwidth]{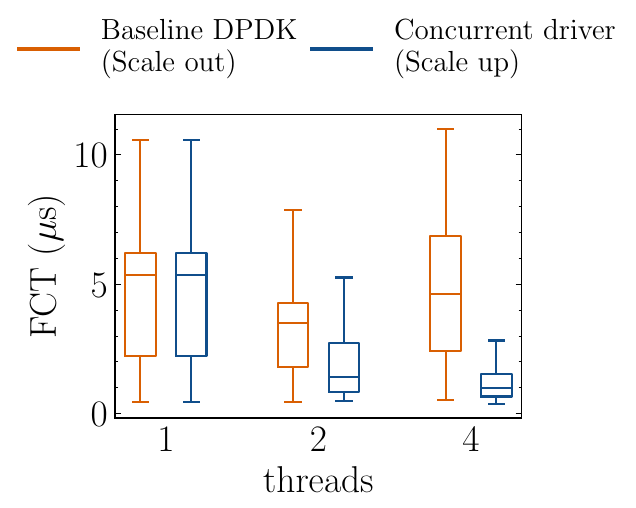}
\subcaption{128 flows}\label{fig:tcp_10k_128}
\end{minipage}
\caption{Flow Completion Time with 10KB payload}
\end{figure*}

We now investigate how a \texttt{l3fwd} router equipped with our multithreaded driver can impact real-world use cases using TCP connections to exchange data. More specifically, the router is connected to a client and a server and forwards traffic between the two. The router is deployed in two different scenarios, scale-out (like the state of the art) and scale-up (our solution). Tests are executed with 1, 2 or 4 threads per NIC: as explained, in the scale-out option each thread has its own Rx queue, while in the scale-up option one queue is shared among all the threads (see Figure \ref{fig:scale}). Our goal is to investigate, in different test scenarios, whether our approach can improve the end-to-end latency or can cause performance degradation because of the retransmissions caused by reordering. We used the standard Linux TCP CUBIC congestion control.\\
We focus on three different test scenarios: one massive flow (High-Performance-Computing-style), many medium flows (ordinary client connections), and many small flows (data-center RPC-style connections). The first case is not compared to the scale-out policy since a unique flow would involve only one Rx queue and thread because of RSS; therefore, there is no chance to distribute the load among cores.\\
Latencies are calculated by retrieving the OS timestamp right before the connection setup (\texttt{connect} syscall) and right after the connection teardown (\texttt{close} syscall).\\
\textbf{Single Huge Flow:}
We test our approach with two different flow sizes, namely 1GB and 10GB. This is expected to be the worst case for our COREC driver since any packet reordering occurs within the single flow being delivered, and thus directly impacts the TCP transmission control protocol. Results are shown in Table \ref{tab:tcphugeflows}. As expected, our approach causes performance degradation in the Flow Completion Time (FCT), owing to the increase in the TCP retransmissions, which are a direct consequence of packet reordering and are also exacerbated when the whole bandwidth of the 10Gbps link is assigned to a unique flow. Still, performance degradation is marginal, with an increase in the flow completion time of 2.3\% in the case of 10GB flow when moving from 1 thread to 4 threads, and even less than 1\% in the 1GB experiment.\\
\begin{table}[]
\tabcolsep=0.11cm
\footnotesize
\begin{tabular}{|l|cc|cc|}
\hline
mode & \multicolumn{2}{|c|}{\bf FCT (s)} & \multicolumn{2}{c|}{\bf Retransmissions} \\ \hline
 & \bf1GB & \bf10GB & \bf1GB & \bf10GB \\ \hline
\textbf{COREC 1 core}  & 0.91625 & 9.13267 & 6.375 & 906.25             \\ \hline
\textbf{COREC 2 cores} & 0.91853 & 9.2238 & 626 & 4073.88              \\ \hline
\textbf{COREC 4 cores} & 0.92067 & 9.35045 & 1071.25	& 5042             \\ \hline
\end{tabular}
\caption{Latency and \# retransmissions for huge flows}
\label{tab:tcphugeflows}
\end{table}
\noindent\textbf{Medium and small flows:} Tests are executed with a 100 KB and 10KB payload per connection. 
We first comment on the results for the 10KB connections
shown in Figure \ref{fig:tcp_10k_64} for 64 flows and Figure \ref{fig:tcp_10k_128} for 128 flows. We can clearly see that our approach brings significant benefits both in mean and tail latency by exploiting its work-conserving capabilities. On the other side, the scale-out case does not scale so well to multiple cores, confirming the theory of poor many-core scaling described in \cite{golestani2019software}.\\
While in this case we can expect that reordering is very unlikely because of the short payload, we now try to increase the payload to 100KB, as this case focuses on a more significant load of $\sim$70 packets per flow. Results are presented in Figures \ref{fig:tcp_100k_64} and \ref{fig:tcp_100k_128}, and they clearly show that our driver still improves the FCT, although in a less evident way.

\begin{figure*}[]
\centering
\begin{minipage}{0.4\textwidth}
  \centering
\includegraphics[width=\textwidth]{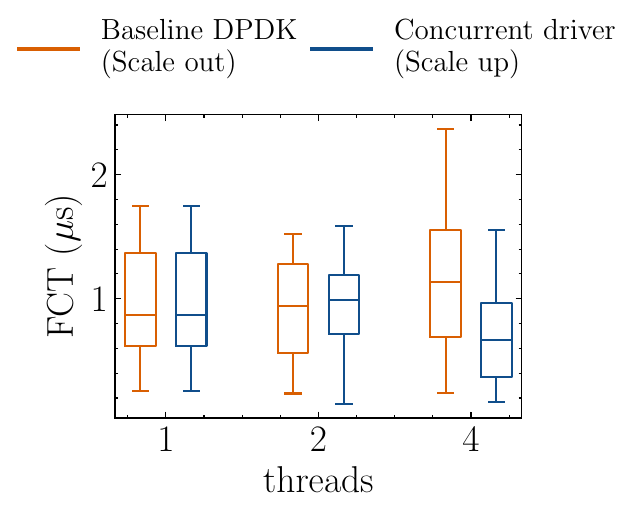}
\subcaption{64 flows}\label{fig:tcp_1k_64}
\end{minipage}
\begin{minipage}{0.4\textwidth}
  \centering
\includegraphics[width=\textwidth]{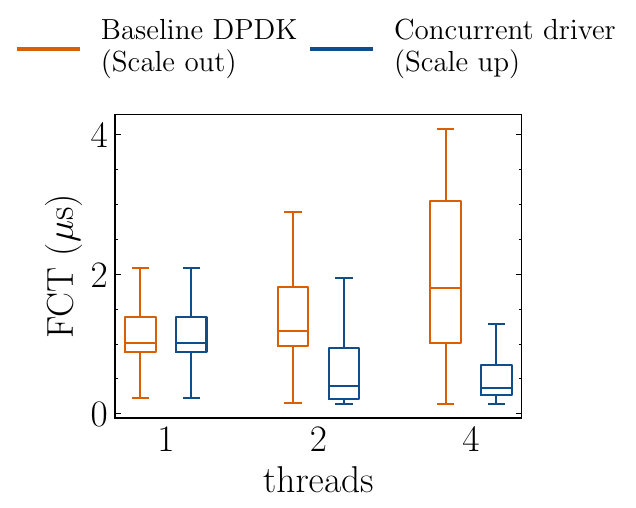}
\subcaption{128 flows}\label{fig:tcp_1k_128}
\end{minipage}
\caption{Flow Completion Time with 1KB payload} \label{fig:tcp_medium_short}
\end{figure*}

\noindent\textbf{One-packet flows:}
It is interesting to also test our driver with one-packet flows (1KB payload). This can be considered as a best case benchmark for TCP traffic,  since, in this case, there is no possibility of re-sequencing the packets at the receiver since only one packet containing TCP payload is sent. Furthermore, these flows are particularly interesting since a significant portion of Data Center flows, especially RPC flows, are restricted to a single packet \cite{homa, dagger}. Also, in this case, we test the FCT time for 64 and 128 TCP parallel flows, and results are shown in Figures \ref{fig:tcp_1k_64} and \ref{fig:tcp_1k_128}. We can clearly see the benefits of our multithreaded driver as the number of flows increases.

\hfill
\hfill

\section{Related Work}\label{sec:rw}
To the best of our knowledge, what we have presented in this article is the first concurrent and non-blocking network driver for a single-queue. This novelty makes it  a complementary solution with respect to what already exists in the literature. 

As a matter of fact, modern network drivers are typically seen as a black box component by researchers, mainly because (i) they are usually developed in the industry with poor documentation \cite{pirelli2020simpler}, and (ii) modern frameworks abstract the details to the programmer for simplicity reasons, thus simply exposing a function to receive/transmit packets. However, in recent years, some works have tried to explain and improve what happens under the hood of a network driver. Ixy \cite{emmerich2019user} is a simple implementation of the \texttt{ixgbe} network driver, with simplicity and educational goals. TinyNF \cite{pirelli2020simpler} simplifies the packet handling for the same driver, showing increased performances and more simplicity at the cost of some flexibility. The work by Emmerich et al. \cite{emmerich2019case} proposes writing network drivers in higher-level languages than common C/C++ implementations to ensure memory safety and reduce bugs. CleanQ \cite{haecki2019cleanq} is a reusable formalization of a driver's descriptor ring with security and portability motivations. 
Also, a few recent works propose low-level code optimizations, either at run time \cite{morpheus} or through a tailored binary file \cite{packetmill}.
Our proposal in this article differentiates from all these solutions since we include in the driver design and implementation concepts inherent to non-blocking management of thread-level concurrency.

The work in  \cite{neugebauer2018understanding} focuses on details related to the 
exchange of packets between the driver and the NIC through PCIe transactions. In particular, 
it shows PCIe's limitations and how it can become a bottleneck for end host networking. Our proposal is complementary to this study since we focus on how the NIC can be managed by a logic integrating non-blocking coordination among threads, independently from limitations related to PCIe.

Metronome, presented in \cite{metronomeconext, metronometon}, uses multiple threads for processing incoming traffic from a same Rx queue.  This is done to cope with reschedule delays that may affect a master thread in charge of processing the flow on a given queue. However, in this solution there is no usage of non-blocking algorithms, rather the threads are synchronized via locking and critical sections.

ShRing \cite{shring} exploits RDMA features like Completion Queues in order to share a single Rx queue among multiple CPUs. The objective is the one of reducing the memory footprint and improving cache effectiveness. However, this solution is still subject to software level synchronization while managing the shared queue.


When processing incoming packets, the NIC and the CPU can interact in two ways; either the NIC informs the CPU of one (or more) packet arrival(s) through an Interrupt ReQuest (IRQ), or the CPU polls the NIC \cite{dpdk,netmap,pfq,zerocopy,snabb}, waiting for new packets. While the former is known to be less invasive in terms of CPU consumption, the latter is known for its better performance \cite{gallenmuller2015comparison} because it avoids the per-packet IRQ overhead. NAPI polling \cite{napi} is an optimization of the former, where once an IRQ is received software busy polls on incoming frames until a certain budget (in terms of descriptors or time elapsed) is consumed. 
Our work focuses on DPDK polling drivers mainly for the simplicity of developing new code and testing it in a friendly user-space environment rather than a kernel one. In any case, we underline that our work could be ported to other environments such as XDP \cite{xdp}, or RDMA, where software polls Completion Queues \cite{kalia2016design}.

RSS++ \cite{barbette2019rss++} tries to load balance traffic between multiple cores by dynamically moving TCP flows from one core to another. PacketSprayer \cite{sadok2018case} tries to spray packets belonging to the same connection among different cores while keeping start/end connection packets on the same one; it exploits the Intel Ethernet Flow Director to achieve this. Willmann et. al. \cite{willmann2006evaluation} discuss the parallelization strategies for network stacks, however, their contribution is limited to lock-based solutions, while we focus on non-blocking operations at the driver level.




As networking moves towards 400GbE speeds, a smart and timely use of cache hierarchies becomes fundamental \cite{cachecloud, darkpackets}. Direct Cache Access (DCA) permits the NIC to place DMAed buffers directly in the L3 cache so that software can find a warm cache, with Intel's DDIO \cite{ddio} being the most popular solution \cite{wang2022understanding}. However, several works showed that DDIO is not a panacea at all: Cai et al. \cite{cai2021understanding} \textit{"observe that it suffers from high cache miss rates (49\%) even for a single flow"}, while \cite{farshin2020,farshin2019make} present optimized DDIO versions in terms of latency.
In the context of more NFs sharing the same L3 cache, \cite{tootoonchian2018resq} proposes a limitation on the number of descriptors for avoiding the \textit{leaky DMA} problem, where LLC cache contention can cause the eviction of packets which still have to be processed by the system. The work in \cite{packetorder} shows how delaying and reordering packets can improve cache locality and therefore, performance. Our solution is essentially orthogonal, since our focus is on the reduction of CPU-cycles for processing incoming traffic thanks to the avoidance of active wait (i.e., spin phases) for making multiple threads access the same Rx queue. This reduction comes together with actual parallelism while processing the data flow from a single queue.

In the last years a set of works were published, either presenting a CPU scheduler \cite{caladan, shenango, mcclurensdi22}, user-level thread-management systems \cite{userlevelthreding, qin2018arachne}, a performance-isolation framework \cite{perfiso} or a specific OS \cite{zygos, kaffes2019shinjuku, ix}: the common goal of such solutions is the efficient schedule of tasks across different cores, explicitly targeting interference mitigation and predictable, low latency for user applications. All of these works try to minimize tail latency by focusing on the application level, while our proposal dives deep into one of the datapath components, namely the network driver.

\section{Conclusions}
This paper has addressed the performance limitations associated with tail latency in modern network stacks. By leveraging multiple receive queues, existing network stacks restrict concurrent processing to a single thread per queue.
To overcome this limitation, we have introduced COREC, a groundbreaking implementation of a concurrent and non-blocking single-queue receive driver.

COREC revolutionizes network stack performance by enabling multiple threads to efficiently share a single queue, thereby enhancing workload distribution and promoting a work-conserving policy. Unlike conventional approaches relying on critical sections, COREC employs atomic machine instructions from the Read-Modify-Write (RMW) class to seamlessly coordinate threads accessing the receive queue. These instructions empower threads to access and update memory locations based on customized conditions, such as matching target values selected by each thread.

The implementation of our novel network driver routine in DPDK v21.11 has showcased its versatility and adaptability. We remark that our contribution is not limited to DPDK and can be extended in the future to other frameworks such as the Linux kernel, XDP, or RDMA completion queues.

Extensive evaluation results have demonstrated the non-critical nature and minimal impact of the occasional additional reordering introduced by our approach. Even under the most challenging circumstances, such as a single large TCP flow, the observed performance impairments have remained within the range of a mere 2-3 percent. Conversely, COREC has consistently achieved remarkable and substantial reductions in latency, especially when handling UDP traffic, real-world traffic mixes, and multiple shorter TCP flows.

\bibliography{biblio2}
\end{document}